\documentclass[prd,reprint,showkeys]{revtex4}
\usepackage{graphics}
\usepackage{amsfonts}
\usepackage{mdframed}
\usepackage{amssymb}
\usepackage{color}
\usepackage{amsmath}
\usepackage{graphicx}
\usepackage[font={footnotesize,it}]{caption}

\begin{document}

\title{Existence of \ traversable wormholes in the spherical stellar systems 
}
\author{A. \"{O}vg\"{u}n}
\email{ali.ovgun@emu.edu.tr}
\author{M. Halilsoy}
\email{mustafa.halilsoy@emu.edu.tr}
\affiliation{Physics Department, Eastern Mediterranean University, Famagusta, Northern
Cyprus, Mersin 10, Turkey.}
\date{\today }

\begin{abstract}
Potentiality of the presence of traversable wormholes in the outer/inner
regions of the halos of galaxies, situated on the Navarro-Frenk-White (NFW)
density profile and Universal Rotation Curve (URC) dark matter models have
been investigated recently \cite%
{raham1,raham2,kuf1,kuf11,kuf111,raham22,raham23}. Since this covers our own
galaxy also as a possible home for traversable wormholes it prompts us to
further the subject by considering alternative density distributions. From
this token herein we make use of the Einasto model \cite{eina1,eina2,eina3}
to describe the density profiles for the same purpose. Our choice for the
latter is based on the fact that theoretical dark matter halos produced in
computer simulations are best described by such a profile. For technical
reasons we trim the number of parameters in the Einasto profile to a
possible minimum. Based on such a model it is shown that traversable
wormholes in the outer regions of spiral galaxies are possible while the
inner part regions prohibit such formations.
\end{abstract}

\maketitle

\section{Introduction}

Although Flamm's work on the wormhole physics dates back to 1916, in
connection with the newly found Schwarzschild solution \cite{flamm},
wormhole solutions were firstly considered from physics standpoint by
Einstein and Rosen (ER) in 1935, which is known today as Einstein-Rosen
bridge connecting two identical space-times \cite{einstein}. Then in 1955
Wheeler used "geons" (self-gravitating bundles of electromagnetic fields) by
giving the first diagram of a doubly-connected-space \cite{wheeler}. Wheeler
added the term "wormhole" to the physics literature, however he defined it
at the quantum scale. After that, first traversable wormhole was proposed by
Morris-Thorne in 1988 \cite{morris}. Then Morris, Thorne, and Yurtsever
investigated the requirements of the energy condition for wormholes \cite%
{morrisb}. After while, Visser constructed a technical way to make
thin-shell wormholes which thoroughly surveyed the research landscape as of
1988 \cite{poisson,visser,visser1,visser2}. After this, there are many
papers written to support this idea \cite{ali,halilsoya,halilsoyb,sakalli}.

In 1933 F. Zwicky, was the first astronomer to propose the existence of the
dark matter (DM) after that he observed the motions of the galaxies in the Coma
Cluster, a galaxy cluster roughly 320 million light-years away and nearly 2
light-years across, and found that it moved too rapidly \cite{zwicky}. He
noted that its speed of revolution, which depends on the weight and position
of the objects inside, implied the cluster had much more mass than he could
see. He decided there must be a hidden ingredient, known as dark matter, that caused the motions of these galaxies to be so large. In 1978,
American astronomer Vera Rubin looked at individual galaxies and she
realized thet the outer stars were spinning as quickly as the interior
stars, and sometimes faster \cite{vera}. It is concluded that the galaxies
must be surrounded by a halo of matter which is not seen. Theorists have
postulated many candidates for what DM might be. But the most widely
accepted hypothesis is the weakly interacting massive particles (WIMPs), a
particle that interacts with gravity but not light, hence its invisibility
to us \cite{cui}. Neither WIMPs nor any other particle that could
successfully explain DM exist in the standard model of particle physics, the
theoretical framework that has successfully predicted nearly all the
phenomena in the universe. Thus, the identity of the DM particle remains one
of the outstanding mysteries in particle physics and cosmology. According to
the latest results from the Planck satellite \cite{planck}, a mere 26.8\% DM
percent of the universe which plays a central role in modeling of cosmic
structure, galaxy formation and evolution and on explanations of the
anisotropies observed in the cosmic microwave background (CMB), 4.9\% is
made of ordinary matter which is matter composed of atoms or their
constituents and 68.3\% is dark energy (DE) that tends to accelerate the
expansion of the universe \cite{Garret,cline,gaskins}. Certain types of DM
can actually help the formation of wormholes. For example, dissipative DM in
which DM particles forming galactic halos are presumed to have dissipative
self-interactions can aid the formation of black holes by dissipative
collapse, and subsequent to wormholes. Now, in this types of models one can
consider either DM collapsed to a disk, or a quasi-isothermal profile \cite%
{fan1,fan2,foot,myr,kril}.

Recently, Rahaman et al. have used both the NFW \cite%
{navarro,raham22,raham23} 
\begin{equation}
\rho (r)=\frac{\rho _{s}}{\frac{r}{r_{s}}\left( 1+\frac{r}{r_{s}}\right) ^{2}%
},
\end{equation}%
and URC DM density profiles \cite{salucci} 
\begin{equation}
\rho (r)=\frac{\rho _{0}r_{0}^{3}}{(r+r_{0})(r^{2}+r_{0}^{2})}
\end{equation}%
where $\rho _{0}$, $\rho _{s}$, $r_{0}$ and $r_{s}$ are all constants, to
show the potentiality of the presence of traversable wormholes in the outer
and the inner part, respectively, of the halo.

In this paper, following similar line of thought, we study the Einasto model 
\cite{eina1,eina2,eina3} which is the special function that arranges the
finest total fit to the halo density profiles. It tells how the density $%
\rho $ of a spherical stellar system changes with distance $r$ from its
center \cite{hjorth}: This is expressed by the density profile%
\begin{equation}
\rho (r)\propto e^{-Ar^{\alpha }}  \label{einastoeqn}
\end{equation}%
with constants $A$ and $\alpha $.

The bounded parameter $\alpha $ which is called Einasto index, adjusts the
degree of curvature of the profile \cite{dhar,retana}

\begin{figure}[h]
\includegraphics[width=0.50\textwidth]{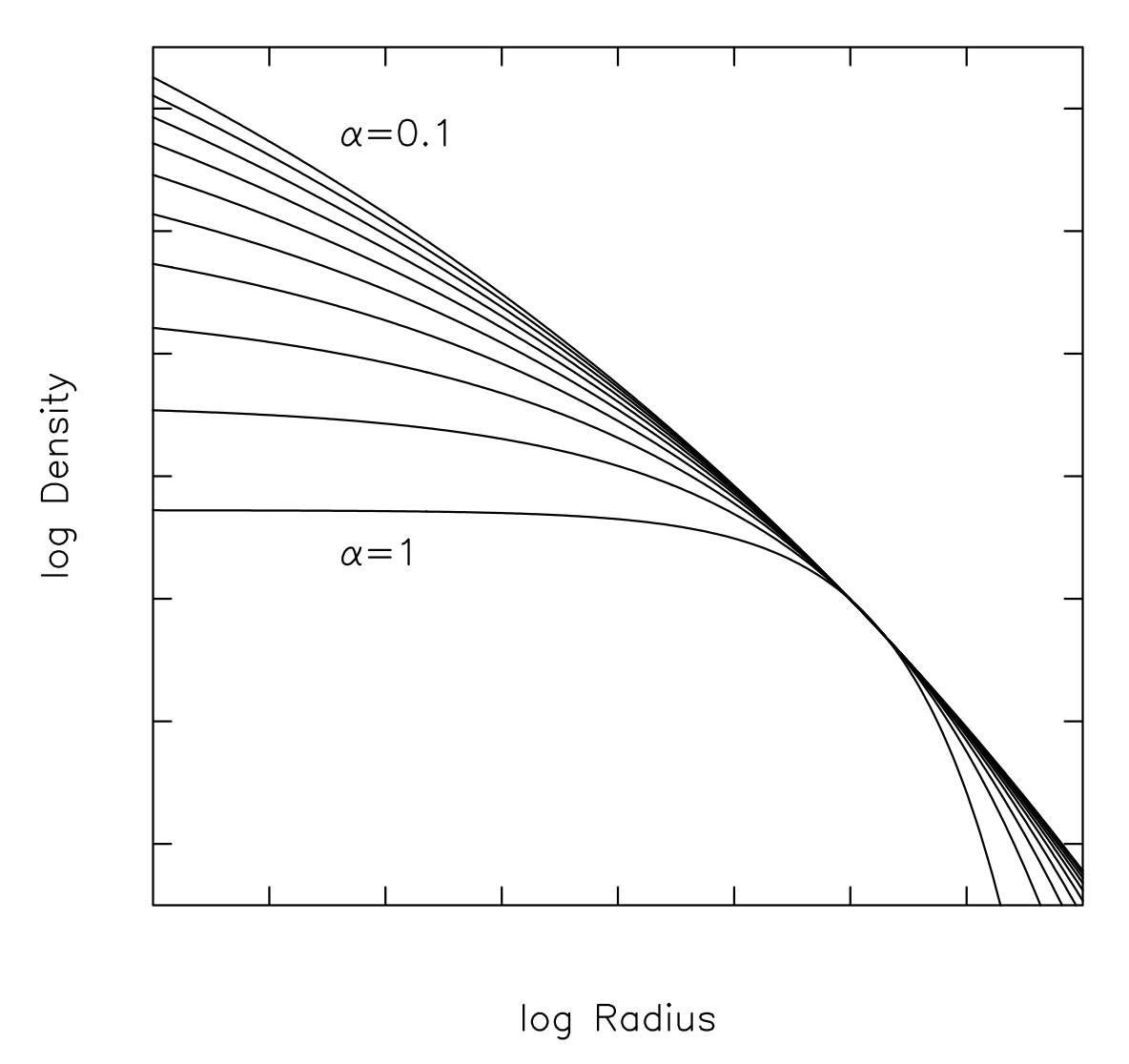}
\caption{Einasto profile in log-log plot (the parameter $\protect\alpha $
controls the degree of curvature of the profile) from Eqn. \protect\ref%
{einastoeqn} .}
\label{einasto1}
\end{figure}

\begin{equation}
\rho (r)=\rho _{0}\exp \left[ -\frac{2}{\alpha }((\frac{r}{r_{0}})^{\alpha
}-1)\right] .  \label{einasto}
\end{equation}

simply named as the Einasto model in the present paper which is plotted in Fig. \ref{einasto1}. For the Milky Way galaxy, we have $r_{0}=9.11$ kpc and $\rho _{0}=5\times 10^{-24}(r_{0}/8.6$~kpc)$^{-1}$gcm$^{-3}$ \cite{maccio,salucci}.

\begin{figure}[h]
\includegraphics[width=0.60\textwidth]{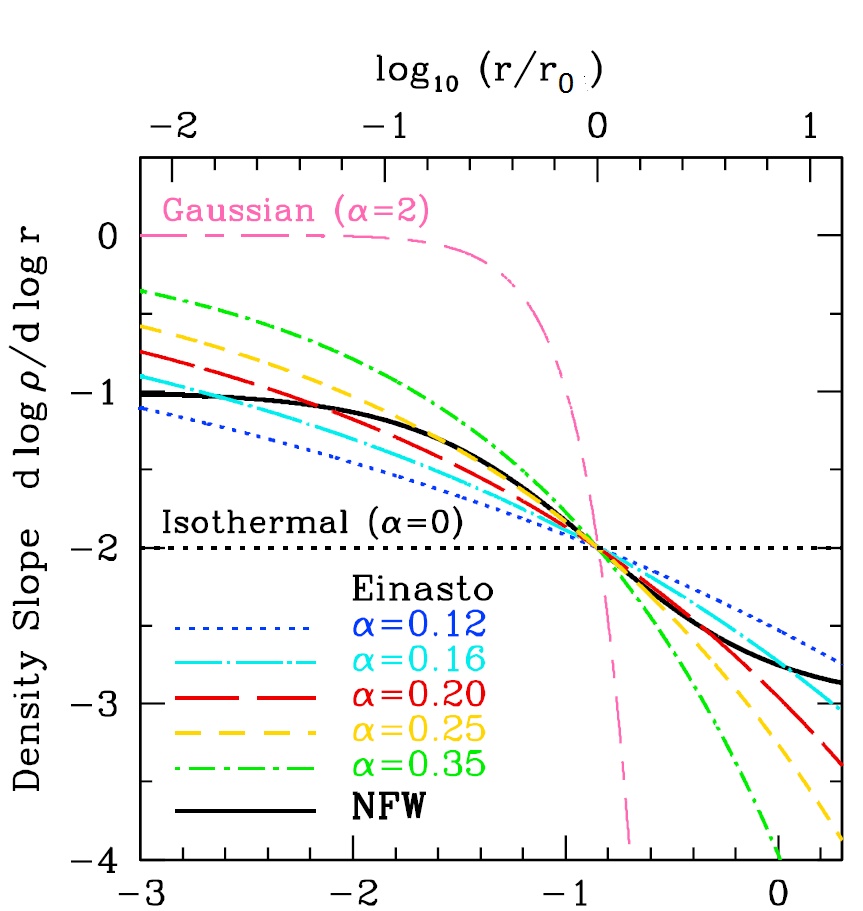}
\caption{Density profiles of Einasto and NFW, density slope (d log $\protect%
\rho $/d log r)), \protect\cite{dutton}}
\label{einastonfw1}
\end{figure}

The slopes of the NFW profile has the range from the inner/outer ( $%
r\rightarrow 0$ / $r\rightarrow \infty $ ) are depicted in Figs. \ref%
{einastonfw1} and \ref{einastonfw2} below. On the other hand, the values are
different for the Einasto profile which are included as limiting cases of
isothermal and Gaussian, ($\alpha =0$, $\alpha =2$, respectively). The
difference between the Einasto and NFW profiles appears at small and large
radial distances. After all, on scales of concern for gas or stellar
dynamics and strong gravitational lensing the Einasto profile with $\alpha
=0.25$ is very akin to the NFW profile, and on scales of concern for weak
gravitational lensing the Einasto profile with $\alpha =0.2$ is quite
similar to the NFW profile. Galaxy mass halos typically have $\alpha =0.16$
as referred in Figs. \ref{einastonfw1} and \ref{einastonfw2} \cite{dutton}.

\begin{figure}[h]
\includegraphics[width=0.60\textwidth]{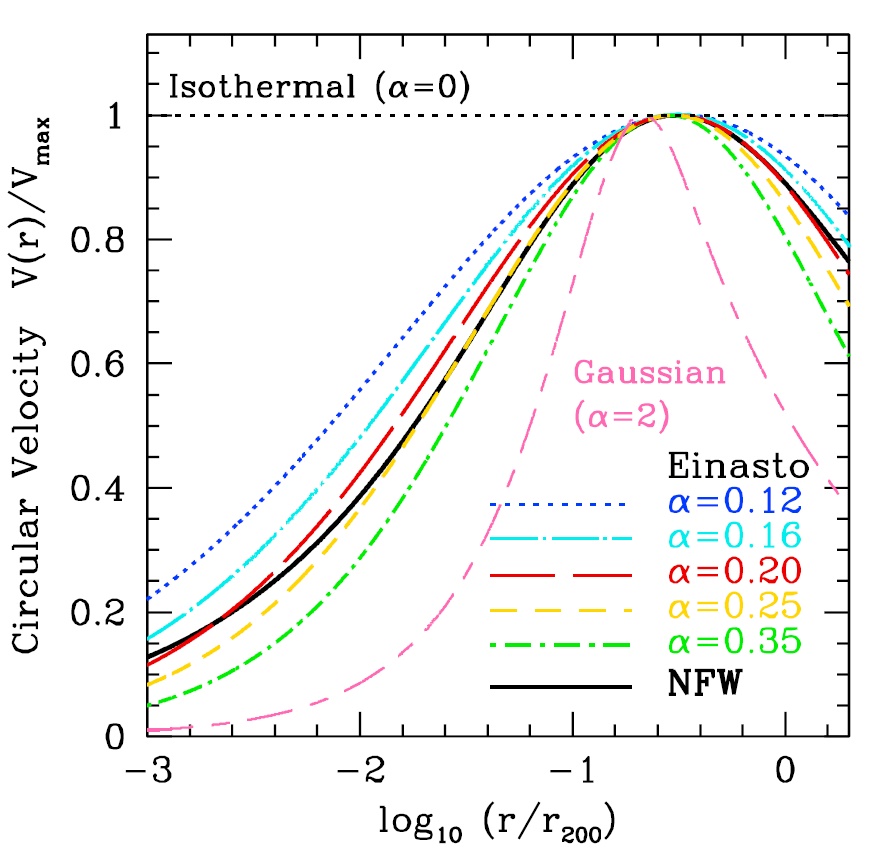}
\caption{Comparison of NFW (solid black lines) and Einasto (colored lines)
profiles. In this example the halo concentration $c_{200} = 7.1$ 
\protect\cite{dutton}}
\label{einastonfw2}
\end{figure}

\begin{figure}[h]
\includegraphics[width=0.75\textwidth]{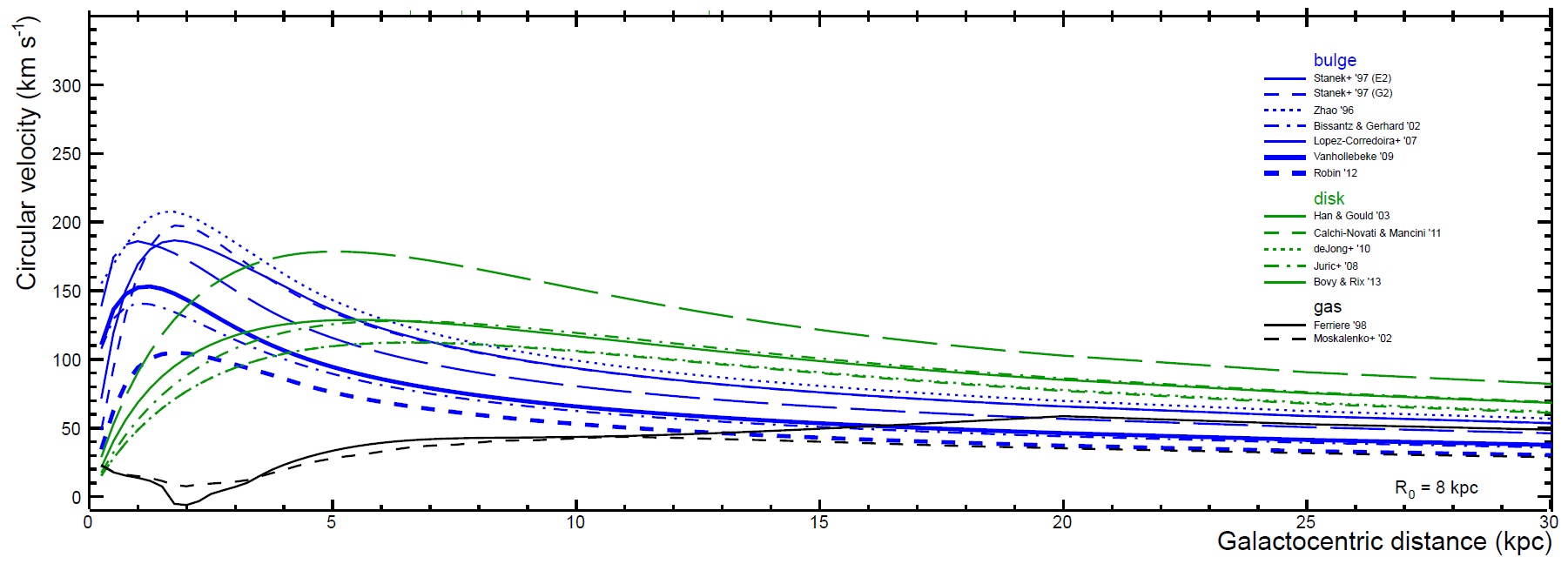}
\caption{Rotational velocity curve for Milkway (the contribution to the
rotation curve as predicted from different models for the stellar bulge
(blue), stellar disk (green) and gas (black). We assume a distance to the
galactic center $r_{0}=8$ kpc \protect\cite{nature})}
\label{milky}
\end{figure}

\begin{figure}[h]
\includegraphics[width=0.75\textwidth]{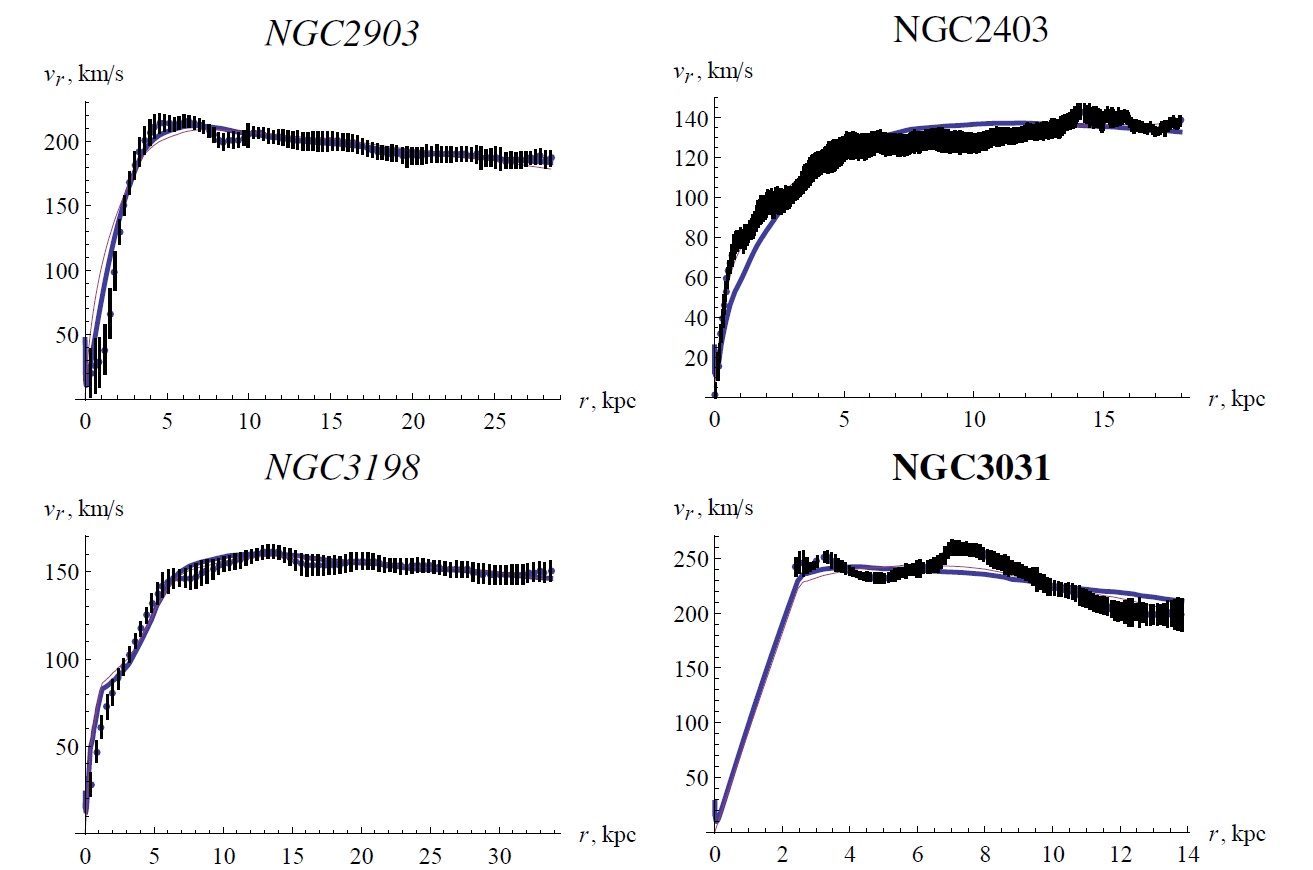}
\caption{Rotational curves v(r) for some galaxies from THINGS survey
together with fits. Blue thick curves show best fits by DM
distributions of semi-degenerate configurations, while magenta thin curves
show Einasto profile best fits \protect\cite{ruffini}}
\label{galaxies}
\end{figure}

Einasto's model has been used in the description of many systems, such as
galaxies, and halos. It is used to show the DM profiles of simulated halos
as well as the NFW profile \cite{bernal,serra,dhar}. The Einasto profile has
a finite (zero) central slope, in contrast to the NFW profile which has a
divergent (infinite) central density. Due to the restricted resolution of
N-body simulations, it is unclear yet which model description fit well to
the central densities of simulated DM halo. The observations of both the
Milky Way and some galaxies may be compatible with the NFW profile for the
DM halo \cite{narikawa,helmi,nature,umetsu,lovell}. There are also
self-interacting dark matter (SIDM) simulations and baryon feedback
simulations \cite{rocha,duffy,chan}. The SIDM is attractive because it
offers a means to lower the central densities of galaxies without destroying
the successes of CDM on large scales. Cosmological simulations that contain
only CDM indicate that DM haloes should be cuspy and with (high)
concentrations that correlate with the collapse time of the halo. One
possible answer is feedback in which the expulsion of gas from galaxies can
result in lower DM densities compared to dissipationless simulations, and
thus bring CDM models in line with observations \cite{rocha,vogel}.
Furthermore baryon feedback has most significant effects on the inner
density profiles which are typically shallower than isothermal and the halo
concentrations tend to be lower than in the absence of baryons. For instance
the strong feedback models reduce the baryon fraction in galaxies by a
factor of 3 relative to the case with no feedback \cite{chan}.

The DM profile of smaller galaxies tend to have flatter
distributions of DM in the central region, known as the cuspy halo problem 
\cite{pato1,pato2}. Our purpose in this paper is to search for possible
traversable wormholes in a spiral galaxy specified by the Einasto density
profile. Wormholes are solutions to Einstein's equations that connect
different universes or tunnels through different parts of the same universe.
Our primary aim is to consider the Milky Way galaxy as a test bed for this
venture. To this end we choose the parameter $r_{0}$ of Milky Way to
coincide with the throat of the traversable wormhole whereas the second
parameter $\alpha $ is in a restricted range. From the Einstein equations we
derive the equations for the density and pressures in terms of the metric
functions and their derivatives. Although our analytical functions are
expressed in terms of the Whittaker functions \cite{whittaker} the detailed
analysis are not imperative. It suffices for us to check the null-energy
condition and satisfaction of the flare-out conditions \cite{morris,morrisb}%
. The same technical handicap prevents us to investigate the stability of
the resulting traversable wormhole that may exist in the Milky Way galaxy.
It is observed that at the central region Einasto profile does not give a
traversable wormhole \ solution because violation of null energy condition
is not satisfied there. But in the outward region we obtain possible
traversable wormholes.

The paper is organized as follows. In Sec. II we present the Einstein
equations for wormhole space-time and we find their solutions under the
Einasto DM profile. In Section III we discuss our results and conclude the
paper.

\section{Traversable Wormholes under the Einasto DM profile}

The spherically symmetric and static wormhole geometry which was proposed by
Morris-Thorne is given by the line element \cite{morris} 
\begin{equation}
ds^{2}=-e^{2f(r)}dt^{2}+\left( 1-\frac{b(r)}{r}\right)
^{-1}dr^{2}+r^{2}(d\theta ^{2}+\sin ^{2}\theta \,d\phi ^{2}),  \label{line}
\end{equation}

where the function $b(r)$ defines the spatial shape function, and $f(r)$
stands for the redshift function. The radial coordinate r ranges from +$%
\infty $ to $b(r_{0})=r_{0}$(minimum radius condition that any wormhole must
satisfy). The minimum value of the r is r$_{0},$ called the throat of the
wormhole. The proper radial distance is given by 
\begin{equation}
l(r)=\pm \int_{r_{0}}^{r}\frac{dr}{\sqrt{\left( 1-\frac{b(r)}{r}\right) }}.
\end{equation}%
Properties of such traversable wormholes are listed below \cite%
{morris,morrisb};

\begin{itemize}
\item Spherically symmetric, static metric,

\item Satisfy the Einstein field equations,

\item No event horizon, i. e. $e^{f(r)}\neq 0,$

\item "Throat" $r=r_{0}$ connects two asymptotically flat space-time
regions, (for Milky Way galaxy $r_{0}=9.11$~kpc)

\item The stress-energy tensor (violating the null-energy condition (NEC)
with $\rho +p_{r}<0$, where $\rho $ is the energy density and $p_{r}$ the
radial pressure),

\item Physically reasonable construction materials, (for Milky Way galaxy $%
\rho _{0}=5\times 10^{-24}(r_{0}/8.6$~kpc)$^{-1}$ g~cm $^{-3}$) 

\item Bearable tidal gravitational forces,

\item Satisfy the flare-out condition at the throat ($b^{\prime }(r_{th})<1$%
), while $b(r)<r$ near the throat,

\item Finite and reasonable crossing time,

\item Stable against perturbations.
\end{itemize}

The Einstein field equation are given by 
\begin{equation}
G_{\nu }^{\mu }=8\pi T_{\nu }^{\mu }
\end{equation}%
where $T_{\nu }^{\mu }$ is the stress-energy tensor, $G_{\nu }^{\mu }$ is
the Einstein tensor with the units in which $c=G=1$. \ Field equations
relate space-time curvature to matter and energy distribution. The non-zero $%
G_{\nu }^{\mu }$ components are 
\begin{equation}
G_{t}^{t}=-\frac{b^{\prime }}{r^{2}},  \label{gtt}
\end{equation}%
\begin{equation}
G_{r}^{r}=\frac{-b}{r^{3}}+2(1-\frac{b}{r})\frac{f^{\prime }}{r},
\end{equation}%
\begin{equation}
G_{\theta }^{\theta }=G_{\phi }^{\phi }=\left( 1-\frac{b}{r}\right) \left[
f^{\prime \prime }++{f^{\prime }}^{2}+\frac{f^{\prime }}{r}-\left( f^{\prime
}+\frac{1}{r}\right) \left\{ \frac{b^{\prime }r-b}{2r(r-b)}\right\} \right] ,
\end{equation}%
in which a prime is $\frac{d}{dr}.$

It is assumed that DM is expressed in the form of general
anisotropic energy-momentum tensor \cite{raham2,null}, as follows 
\begin{equation}
T_{\nu }^{\mu }=(\rho +p_{r})u^{\mu }u_{\nu }+p_{r}g_{\nu }^{\mu
}+(p_{t}-p_{r})\eta ^{\mu }\eta _{\nu },
\end{equation}%
in which $u^{\mu }u_{\mu }=-\frac{1}{2}\eta ^{\mu }\eta _{\mu }=-1$, and
where $p_{t}$ is the transverse pressure, $p_{r}$ is the radial pressure and 
$\rho $ is the energy density. A possible set of $u^{\mu }$ and $\eta ^{\mu }
$ are given by $u^{\mu }=(e^{2f(r)},0,0,0)$ and $\eta ^{\mu }=(0,0,\frac{1}{r%
},\frac{1}{r\sin \theta })$. Let us add that once we restrict ourselves
entirely to a collisionless cold DM the pressure terms should not
be taken into account. Our choice implies accordingly that we aim a more
general anisotropic fluid, at least for our future interest. The only
non-zero components of $T_{\nu }^{\mu }$ are 
\begin{equation}
T_{t}^{t}=-\rho ,  \label{Tr}
\end{equation}%
\begin{equation}
T_{r}^{r}=p_{r},
\end{equation}

\begin{equation}
T_{\theta }^{\theta }=T_{\phi }^{\phi }=p_{t}.  \label{Ttheta}
\end{equation}

\begin{figure}[h]
\includegraphics[width=0.70\textwidth]{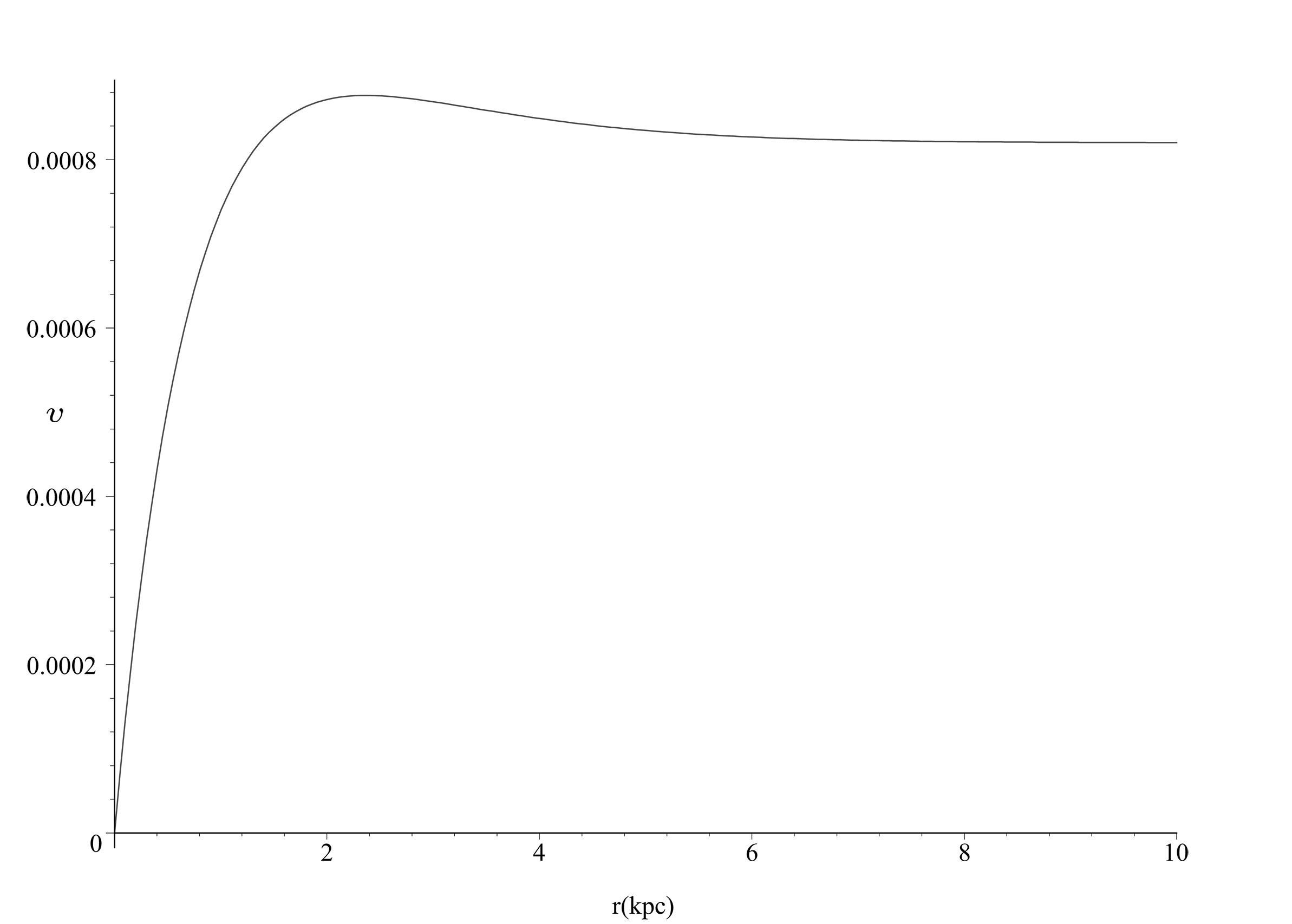}
\caption{Proposed rotational velocity with values of the parameters as $%
k_{1}=k_{2}=1$,$\protect\alpha =0.0006$ and $\protect\beta =0.00082$ 
\protect\cite{raham2} }
\label{velo}
\end{figure}

Furthermore we need to impose some constraints in order to solve Einstein
field equations. One of them is the tangential velocity, (for a
justification of this condition we refer to the book by Chandrasekhar \cite%
{chandra,landau}) 
\begin{equation}
v^{\phi }=\sqrt{rf^{\prime }}  \label{v1}
\end{equation}%
which is responsible to fit the flat rotational curve for the DM.
Other one is proposed by Rahaman et. al \cite{raham2} that the observed
rotational curve profile in \ the DM region is given by%
\begin{equation}
v^{\phi }=\alpha r\exp (-k_{1}r)+\beta \lbrack 1-\exp (-k_{2}r)]  \label{v2}
\end{equation}%
in which $\alpha ,$ $\beta ,$ $k_{1},$ and $k_{2}$ are constant positive
parameters. It is illustrated in Fig.\ref{velo}. With this choice it is
guaranteed that for $r\longrightarrow \infty $ we obtain $v^{\phi }=\beta =$%
constant which is required for flat rotation curves. 

From \ the Eqns.(\ref{v1}) and (\ref{v2}), one obtains the redshift function
as follows%
\begin{eqnarray}
f(r) &=&-\frac{\alpha ^{2}r}{2k_{1}e^{(2k_{1}r)}}-\frac{\alpha ^{2}}{%
4k_{1}^{2}e^{(2k_{1}r)}}-\frac{2\alpha \beta }{k_{1}e^{(k_{1}r)}}+\frac{%
2\alpha \beta e^{(-k_{1}r-k_{2}r)}}{k_{1}+k_{2}}  \notag  \label{E:redshift}
\\
&&+\beta ^{2}\ln (r)+2\beta ^{2}E_{i}(1,k_{2}r)-\beta ^{2}E_{i}(1,2k_{2}r)+D.
\label{fr}
\end{eqnarray}%
with the exponential integral E$_{i}$ \cite{jeffreys} and integration
constant $D$. For large $r$, it is required also that $%
e^{2f(r)}=B_{0}r^{(4v^{\phi })}$\cite{rah3,rah4,chandra}.

\begin{figure}[h]
\includegraphics[width=0.70\textwidth]{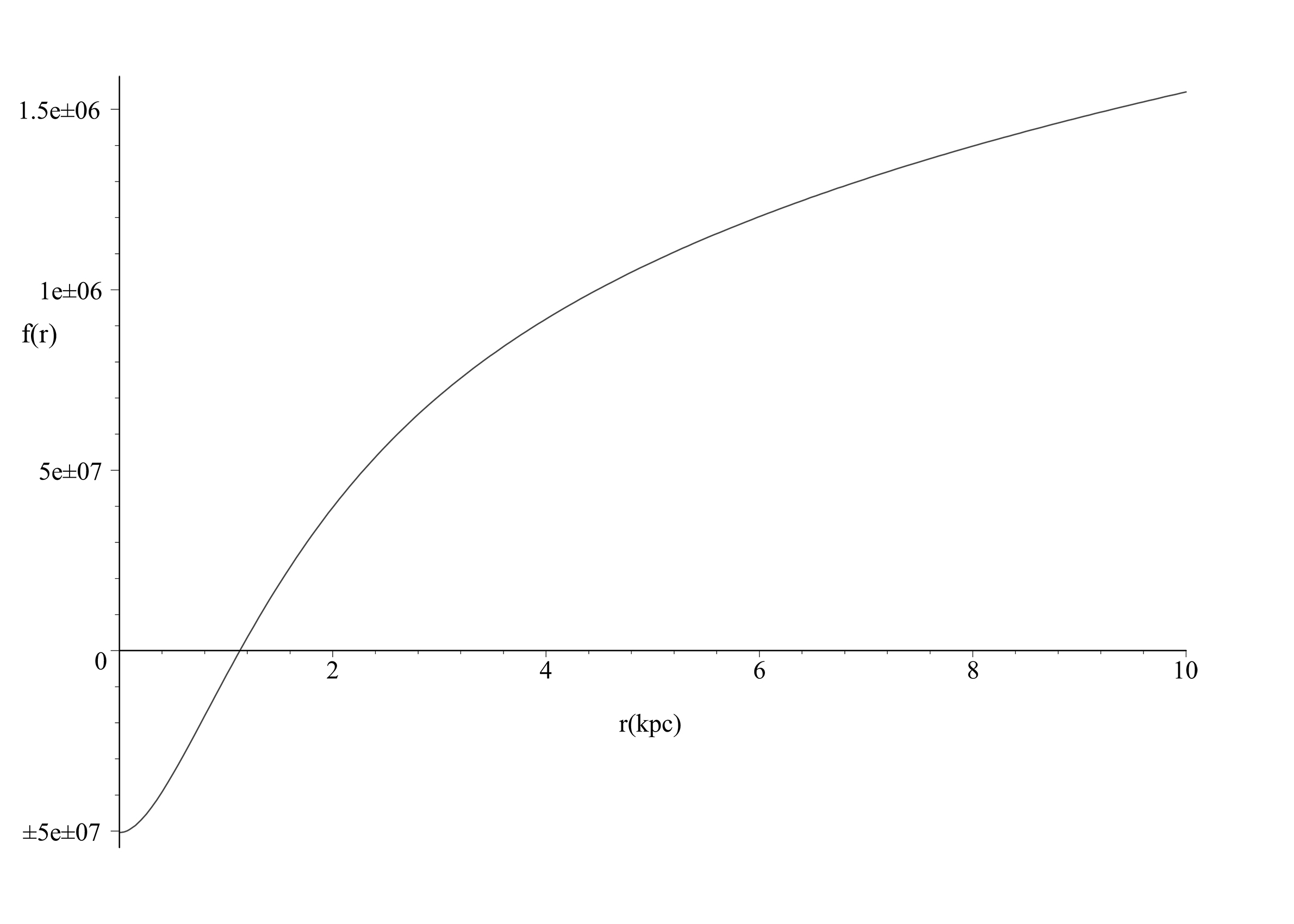}
\caption{f(r) versus r ($k_{1}=k_{2}=1$,$\protect\alpha=0.0006$ and $\protect%
\beta=0.00082$ \protect\cite{raham2} )}
\label{frplot}
\end{figure}

By using the Eqns. (\ref{v1}) and (\ref{v2}) and the Einasto density profile
(Eqn. (\ref{einasto})), the Einstein field equations give us the tedious
form of the shape function%
\begin{equation}
b(r)=\frac{2^{(3-\frac{3}{\alpha })}\pi \rho _{0}}{\alpha }\exp (\frac{2}{%
\alpha })\left( \frac{(r/r_{0})^{\alpha }r^{-\alpha }}{\alpha }\right)
^{-3/\alpha }[\left( \frac{1}{3(\alpha +3)(2\alpha +3)}\right) \alpha
^{3}r^{3}2^{(3/\alpha -1-\left( \alpha +3\right) /2\alpha )}\left( \frac{%
(r/r_{0})^{\alpha }r^{-\alpha }}{\alpha }\right) ^{3/\alpha }  \label{breqn}
\end{equation}%
\begin{equation*}
\times \left( 2(r/r_{0)}\right) ^{\alpha }+\alpha +3)(r/r_{0})^{-\alpha
}\left( \frac{(r/r_{0})^{\alpha }}{\alpha }\right) ^{-(\alpha +3)/2\alpha }
\end{equation*}%
\begin{equation*}
\times \exp (-\frac{(r/r_{0})^{\alpha }}{\alpha })Whittaker M(3/\alpha
-(\alpha +3)/2\alpha ,(\alpha +3)/2\alpha +1/2,\frac{2(r/r_{0})^{\alpha }}{%
\alpha })
\end{equation*}%
\begin{equation*}
+\left( \frac{1}{3(2\alpha +3)}\right) \alpha ^{2}r^{3}2^{(3/\alpha
-1-\left( \alpha +3\right) /2\alpha )}\left( \frac{(r/r_{0})^{\alpha
}r^{-\alpha }}{\alpha }\right) ^{3/\alpha }(\alpha +3)(r/r_{0})^{-\alpha }
\end{equation*}%
\begin{equation*}
\times \left( \frac{(r/r_{0})^{\alpha }}{\alpha }\right) ^{-(\alpha
+3)/2\alpha }\exp (-\frac{(r/r_{0})^{\alpha }}{\alpha })Whittaker M(3/\alpha
-(\alpha +3)/2\alpha +1,(\alpha +3)/2\alpha +1/2,\frac{2(r/r_{0})^{\alpha }}{%
\alpha })]+C_{1}
\end{equation*}

where the integration constant is chosen as (in order that $b(r_{0})=r_{0}$
is satisfied) 
\begin{equation}
C_{1}=\frac{-r_{0}}{6\alpha ^{2}+27\alpha +27}\left( A\pi r_{0}^{2}\exp
(1/\alpha )2^{(3\alpha -3)/2\alpha )}\alpha ^{1+(\alpha +3)/2\alpha }\right)
\label{constant}
\end{equation}

\begin{equation*}
\lbrack \alpha (5+\alpha )Whittaker M(-(\alpha +3)/2\alpha ,(2\alpha
+3)/2\alpha ,2/\alpha )
\end{equation*}%
\begin{equation*}
+(3+\alpha )Whittaker M((\alpha +3)/2\alpha ,(2\alpha +3)/2\alpha ,2/\alpha
)-6\alpha -9].
\end{equation*}

Note that in this expression WhittakerM stands for the Whittaker function 
\cite{whittaker}. The second order differential equation satisfied by $b(r)$
in our present analysis is given by 
\begin{equation}
b^{\prime \prime }=\frac{2b^{\prime }}{r}\left[ 1-\left( \frac{r}{r_{0}}%
\right) ^{\alpha }\right] 
\end{equation}%
which follows from Eq.(\ref{Tr}). The redshift function $f(r)$ versus radial
coordinate $r$ is plotted in Fig.\ref{frplot}.

\begin{figure}[h]
\includegraphics[width=0.70\textwidth]{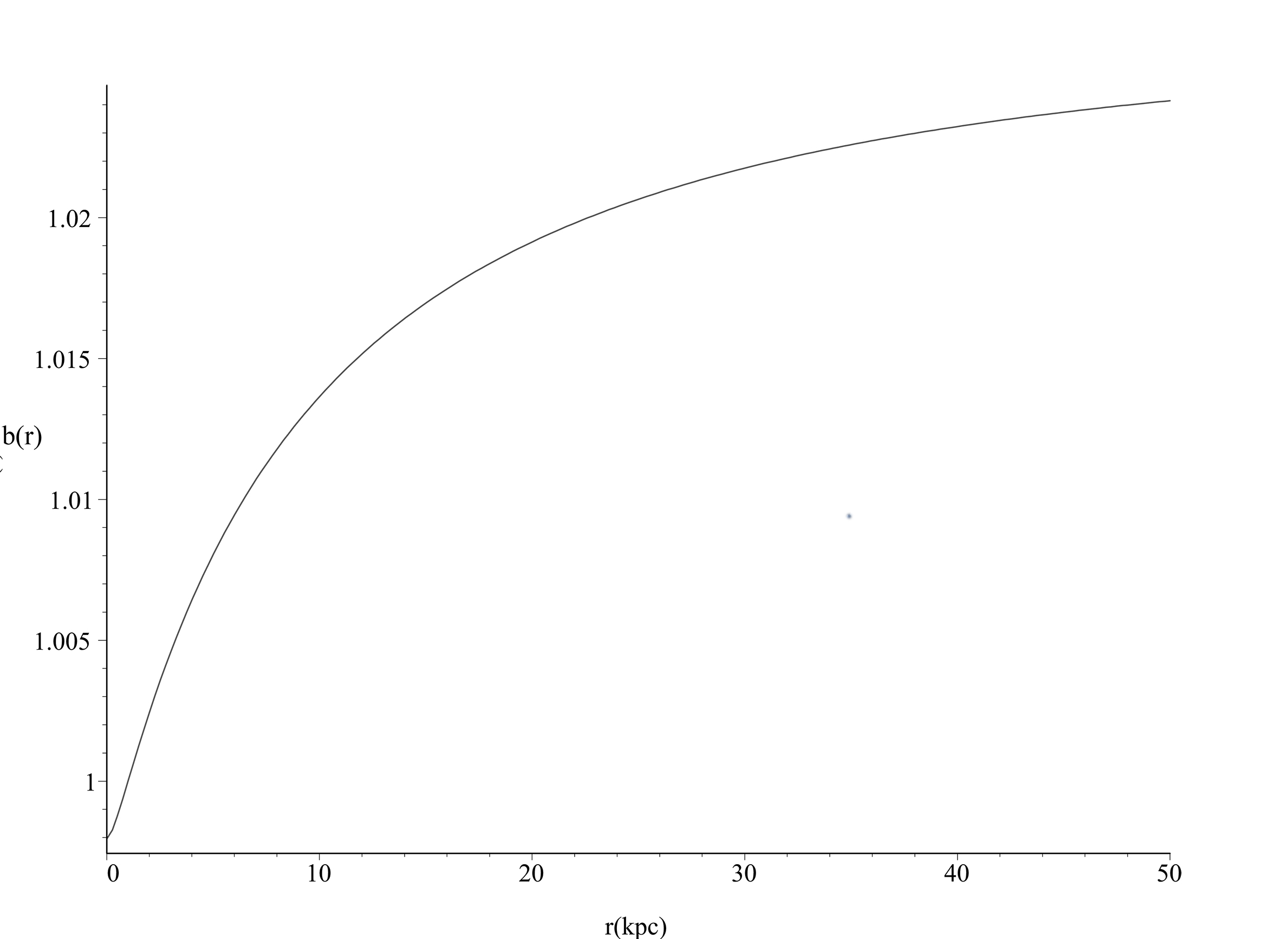}
\caption{The shape function b versus radial coordinate r with the parameters 
$r_{0}=1$ and $\protect\rho _{0}=0.0001$ }
\label{ab}
\end{figure}

\begin{figure}[h]
\includegraphics[width=0.70\textwidth]{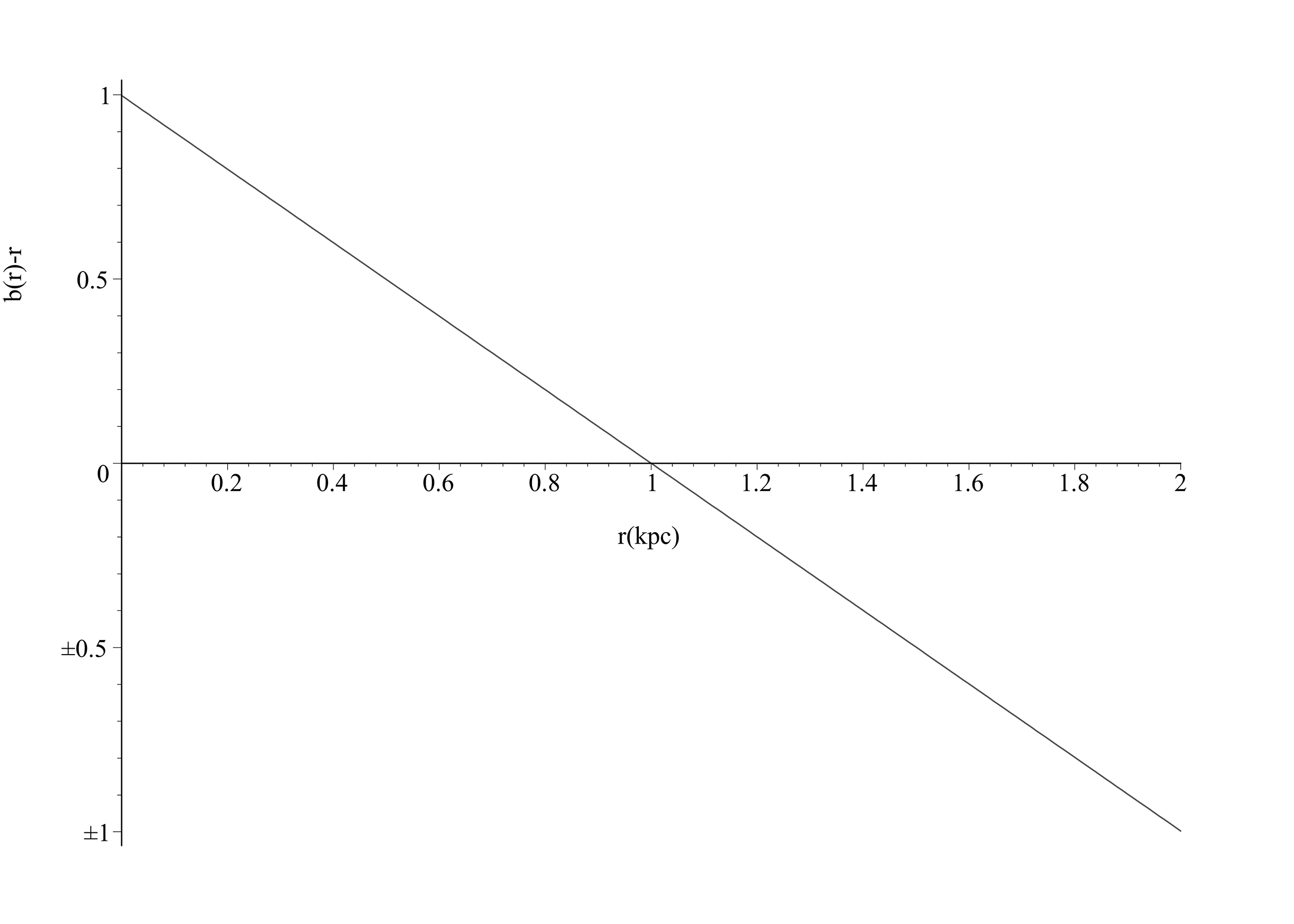}
\caption{The radius of the throat is obtained where b(r)-r cuts r axis or
the values $r_{0}=1$ and $\protect\rho _{0}=0.0001$ }
\label{abr}
\end{figure}

\begin{figure}[h]
\centering
\includegraphics[width=0.70\textwidth]{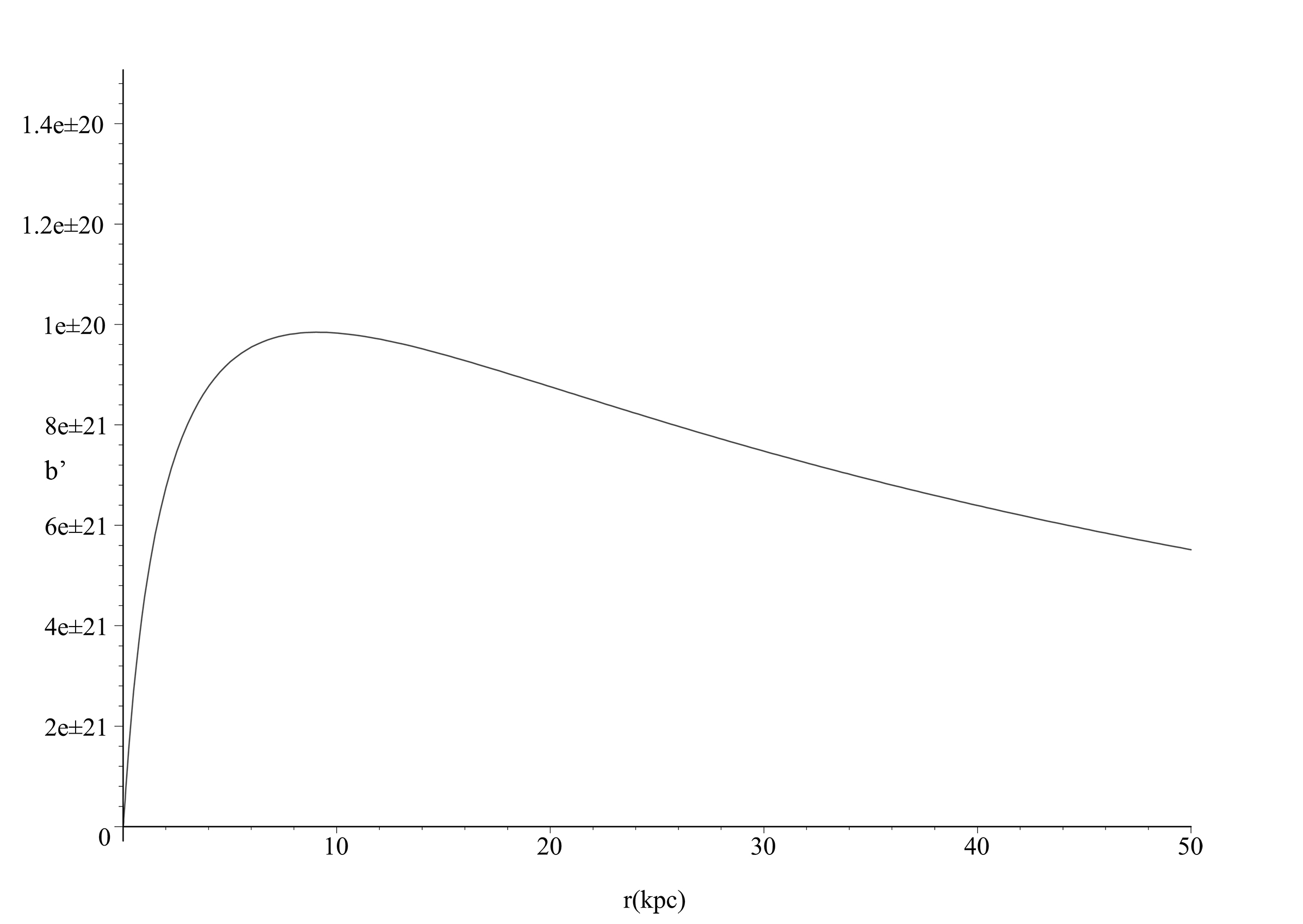}
\caption{$b^{\prime }$ versus r for Milky Way ($r_{0}=9.11$~kpc and $\protect%
\rho _{0}=5\times 10^{-24}(r_{0}/8.6$~kpc)$^{-1}$ g~cm $^{-3}$ \protect\cite%
{salucci,maccio,salucci2})}
\label{b1milky}
\end{figure}

\begin{figure}[h]
\includegraphics[width=0.70\textwidth]{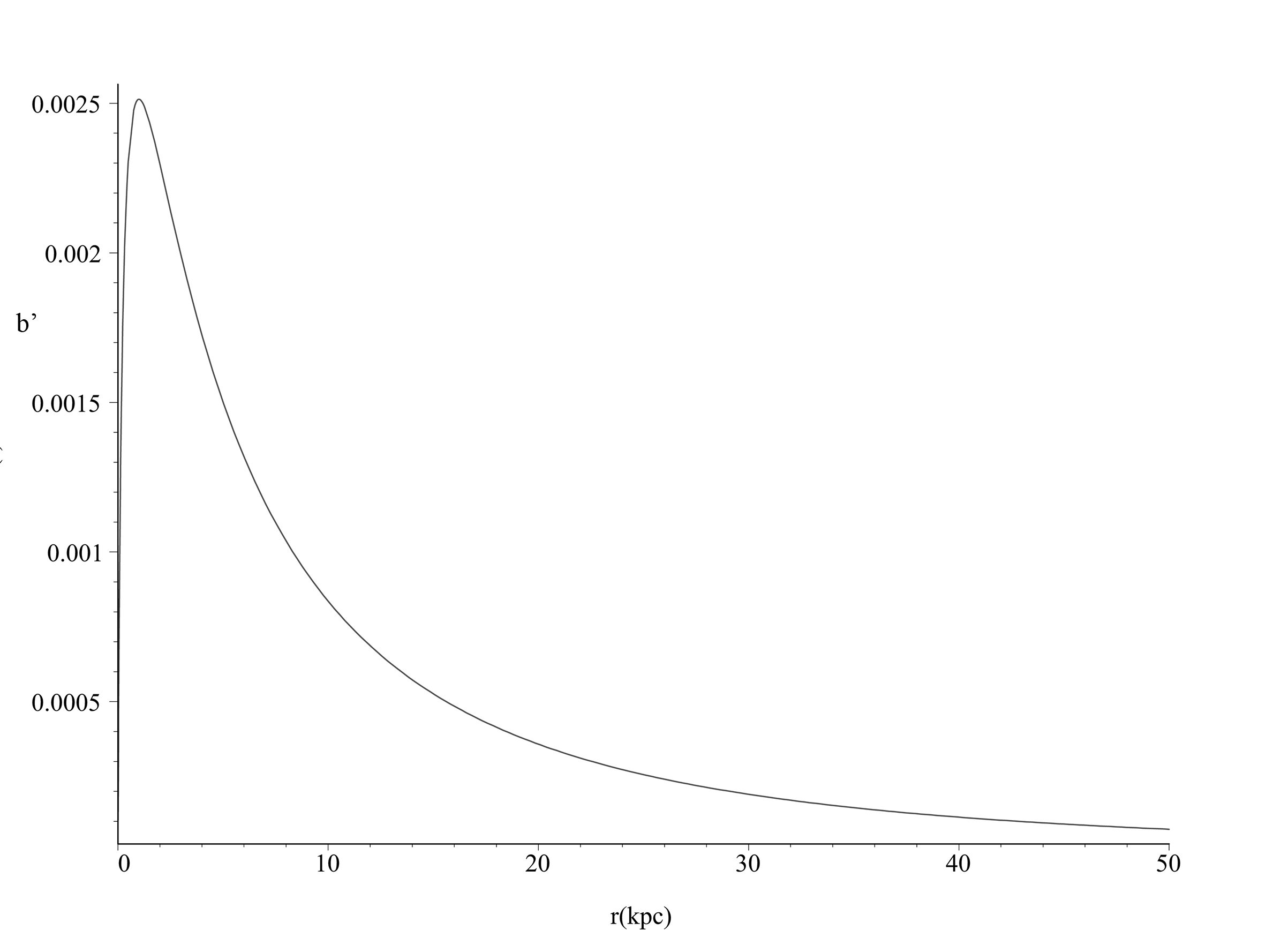}
\caption{$b^{\prime }$ versus r ($r_{0}=1$ and $\protect\rho _{0}=0.0001$) }
\label{ab1}
\end{figure}

It can be checked that the \ wormhole throat condition of $b(r_{0})=r_{0}$
is satisfied. It is shown in Figs. \ref{b1milky} and \ref{ab1}. For the
Milky Way galaxy, we have $r_{0}=9.11$~kpc and $\rho _{0}=5\times
10^{-24}(r_{0}/8.6$~kpc)$^{-1}$ g~cm $^{-3}$ \cite{salucci,maccio,salucci2}
in order that $r_{0}$ is greater than the horizon of the supermassive black
hole. This is necessary for the wormhole to be traversable. The choice of $%
r_{0}$ as throat radius can be justified from the fact that $r_{0}$ plays
the role of the core radius for the galaxy. For this \ reason we rely on Eq.(%
\ref{einasto}) since for $r>r_{0}$ there is a turning point for the density
distribution. We admit, however, that the exact choice of $r_{0}$ as the
throat radius, unless a better choice is made, is open to criticism.
Furthermore, the flare-out condition ($b^{\prime }<1$) 
\begin{equation}
b^{\prime }=8\pi \rho r^{2}=8\pi r^{2}\rho _{0}\exp \left[ -\frac{2}{\alpha }%
((\frac{r}{r_{0}})^{\alpha }-1)\right]   \label{flr}
\end{equation}%
is also satisfied for the Milky Way galaxy $b^{\prime }(r_{0})=0.98\times
10^{-20}<1$ (Fig.\ref{b1milky}) and also all spherical stellar systems as
shown in Fig. \ref{ab1}. In addition, both the star-count and kinematic data
of the Milky Way stellar halo are well-represented by an Einasto profile
with index $\alpha \approx 0.5$ and effective radius $\approx 20$ kpc, if
the dark halo has a flat rotation curve (Figs. \ref{milky} and \ref{galaxies}
)\cite{evans} .

Finally, for the sake of completeness the radial and lateral pressures are
calculated easily from Eqs.(\ref{gtt}-\ref{Ttheta}) as

\begin{equation}
p_{r}=\frac{1}{8\pi }\left( \frac{-b}{r^{3}}+2(1-\frac{b}{r})\frac{f^{\prime
}}{r}\right) ,
\end{equation}%
\begin{equation}
p_{t}=\frac{1}{8\pi }\{\left( 1-\frac{b}{r}\right) \left[ f^{\prime \prime }+%
{f^{\prime }}^{2}+\frac{f^{\prime }}{r}-\left( f^{\prime }+\frac{1}{r}%
\right) \left( \frac{b^{\prime }r-b}{2r(r-b)}\right) \right] \}
\end{equation}

in which $f(r)$ and $b(r)$ are to be substituted from the solution (Eqns. \ref%
{fr}- \ref{constant}). 

Note that the null energy condition ($\rho +p_{r}<0$) is violated so that
the flare-out condition for a wormhole is satisfied according to Fig. \ref%
{null}. Let us note also that those pressures have not been used in the
present study but it can be anticipated that for a future stability analysis
they will be needed.

\begin{figure}[h]
\includegraphics[width=0.70\textwidth]{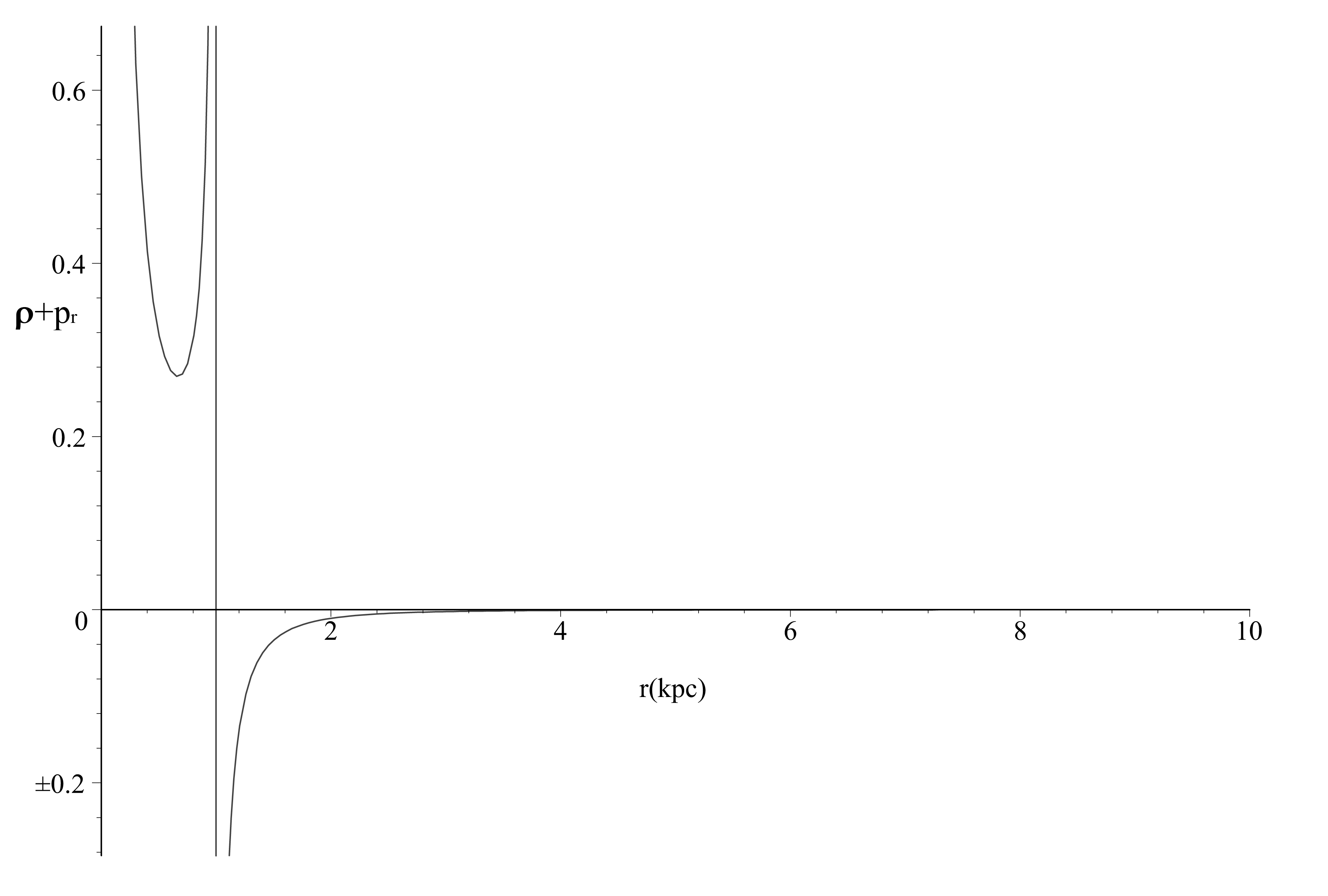}
\caption{Null energy condition ($\protect\rho +p_{r}<0$) for the values of
parameters $r_{0}=1$ and $\protect\rho _{0}=0.0001$ }
\label{null}
\end{figure}

\section{Conclusion}

Existence of traversable wormholes is an important problem in physics both
at micro and macro scales. There is no doubt that together with black holes
wormholes constitute the most interesting objects in our entire universe. On
the one hand collision of ultra-energetic particles may create these objects
while at the other extreme they form naturally in galactic systems of cosmic
dynamics. Once formed, in order to survive traversable wormholes must
satisfy certain criteria of stability, otherwise they will decay instantly
into more stable objects. Besides stability energy conditions are also of
prime importance concerning traversable wormholes. This demands negative
energy density which is not available in classical physics although there
are rooms for it in the quantum domain. Let us add also that a recent trend
is to modify the geometry of throats and obtain thin-shell wormholes \cite%
{ali,sakalli} supported by positive total energy, for this see \cite%
{halilsoya,halilsoyb} and references cited therein. However, the minimum
requirement of violationg null-energy condition (NEC) provides that $\rho
+p_{r}<0$, without specifying the (-) condition for the energy density. As a
matter of fact NEC provides us a jumping board into the realm of traversable
wormholes. Stability, on the other hand, admittedly and for completely
technical reasons remains open in our present study. This item may be
considered as a separate subject in our future analysis.

This work is motivated mainly by Ref. \cite{raham2}, which discusses the
possible formation/existence of traversable wormholes in our own galaxy,
Milky Way. Their conclusion that traversable wormholes may exist both at the
inner and outer parts of our spiral galaxy has been revised by using a
different density distribution. Namely, we adapt the Einasto density profile 
\cite{eina1,eina2} which differs much from the Navarro- Frenk- White (NFW)
profile employed in \cite{raham2,navarro}. Our analysis suggests accordingly
that at the central region since the NEC is not violated, we do not get any
traversable wormholes. This explains also the choice of the throat outside
of the central region. Yet at the outer regions, the NEC is violated so that
the flare-out condition for traversable wormhole is satisfied and we expect
formation of traversable wormholes. This result is not specific to Milky Way
alone but is valid for any spiral galaxy once it obeys a mass distribution
given by Einosto profile. It is worthwhile therefore to check other mass
distributions and see the resulting consequences. In conclusion we wish
to comment that the DM lurking in the innermost part of the Milky
Way \cite{nature}, may support the existence of traversable wormholes that
contains our solar system as well.

\section{Acknowledgment}

We would like to thank the anonymous reviewers for their useful comments and
suggestions which helped us to improve the  paper.

\end{document}